\documentclass[11pt,a4paper]{article}
\usepackage[hyperref]{acl2018}
\usepackage{times}
\usepackage{latexsym}

\usepackage{url}
\usepackage{graphicx}
\aclfinalcopy

\title{Social Status and Communication Behavior in an Evolving Social Network}

\author{Sahand Akbari \\
  Department of Computer Science\\
  University of Toronto\\
  {\tt sahand@cs.toronto.edu} \\}

\date{}

\begin{document}
\maketitle
\begin{abstract}
    The degree to which individuals can exert influence on propagation of information and opinion dynamics in online communities is highly dependent on their social status. Therefore, there is a high demand for identifying influential users in a community by predicting their social position in that community. Moreover, understanding how people with various social status behave, can shed light on the dynamics of interaction in social networks. In this paper, I study an evolving online  social network originated from an online community for university students and I tackle the problem of forecasting users' social status, represented as their PageRank, based on frequency of  recurring temporal sequences of observed behavior, i.e. behavioral motifs. I show that individuals with different values of PageRank exhibit different behavior even in early weeks since the online community's inception and it is possible to forecast future PageRank values given frequency of behavioral motifs with high accuracy.  
\end{abstract}

\subsection *{Author Keywords}
Behavioral motifs, Sequence Analysis, User Behavior, Social Status, Network Evolution, Page Rank

\section{Introduction}
Nowadays, online communities are prevalent and user participation is an inseparable part of internet services and applications in variety of shapes and forms. However, understanding user behavior in online communities has proven to be a challenging task. Many research studies have investigated the methods and applications of understanding users' behavior both in online and offline spaces \cite{Eagle09eigenbehaviors:identifying}. Improving user experience, customized user interfaces, advertisement are a few of the applications of modelling and analysis of user behavior in online communities \cite{Benevenuto:2009:CUB:1644893.1644900} \cite{6588663}. Many studies explore characterizing different classes of users based on their observed behavior. It has been shown that user behavior is indicative of user demographics \cite{Wang:2016:UCC:2858036.2858107} \cite{Hu:2007:DPB:1242572.1242594}\cite{Zhong:2015:YYG:2684822.2685287}, social circles \cite{Mislove:2010:YYK:1718487.1718519}, fake vs. real user accounts in online communities \cite{DBLP:journals/corr/VarolFDMF17} \cite{6280553}, depression \cite{predicting-depression-via-social-media},performance in MOOCs \cite{Anderson12}, and user location \cite{Mahmud:2014:HLI:2648782.2528548} to name a few.  Furthermore, understanding users’ behavior can provide insights on their status and position in the social networks they form \cite{Riquelme:2016:MUI:2978438.2978529} \cite{5484779}.
\newline
In this paper, I  study the behavior of a group of university students that are members of an evolving online communication network. I model users’ behavior as temporal sequences that consist of events. Events considered in the data set are joining the network and sending and receiving messages. Then, I find recurring patterns of behavior, i.e. behavioral motifs, and investigate similarity and differences between members of this online community in terms of the frequency of observed recurring behavioral motifs. I track user behavior and social position, calculated as PageRank, through time and forecast users' social position given cumulative information on user behavior in different weeks since the network inception. It is shown that an individual's social position in the network can be forecasted with relatively high accuracy given behavioral information which suggest high degree of correlation between behavior and future social position. Thus, I investigate the following research question: \textbf{Can temporal behavioral sequence of an individual predict or forecast their social position in an online community?} which is a operationalization of the following motivating research question: \textbf{Do people with different levels of social position in a community distinguish themselves in terms of their behavioral attributes?} Note that the focus here is to underline the value of analyzing temporal sequences of user behavior rather than measuring how active one is an online community. In fact, I will show that while magnitude of activity has correlation with social position, investigating behavior in a fine grained matter can give us more information on social position, measured in terms of prediction accuracy of PageRank values.
To the best of my knowledge, no previous work has explored the possibility of forecasting future social positions measured by PageRank of members of an online community based on observations of fine grained sequences of behavior in terms of frequent behavioral motifs.
\section{Related Work}
This work sits in the intersection of sequence analysis, user behavior, network evolution, and user influence and centrality. While this brief literature review can't do justice to all these research fields, I highlight the most relevant work to my research.

\subsection{User Behavior}
Traditional studies of user behavior in online social communities has largely employed two approaches: Some research studies aggregate sequence logs and create snapshots of the network and investigate various phenomena by using the aggregated information. Other studies define a tie between users based on statistics from the sequence data and use social network analysis to answer a specific structural research questions \cite{keegan} \cite{VanDerAalst:2005:DSN:1107360.1107389} \cite{jurgens}.
here are some studies that take account of temporal features in the data, however they do not specifically investigate sequence of action\cite{Eagle09eigenbehaviors:identifying}\cite{wang}. However,  I express the behavior of an individual in terms of the frequent temporal sub sequence of actions they perform.
\newline
There has been a lot of recent interest in analyzing temporal sequences of behavior. Wang et al. \cite{Wang:2016:UCC:2858036.2858107} built an unsupervised system to identify the dominating user behaviors in online social networks. Using event log data of users actions in two social networks they created sequences of temporal behavior and leveraged iterative feature pruning to hierarchically cluster similar users together. Each parent cluster is further divided to different children clusters by using only the discriminative features(behavioral motifs) and excluding common features from the parent cluster. They also designed a visualization tool that allows users to analyze the clusters in a exploratory manner. A hierarchical analysis of similar users is a strength of their work as it provides a categorization of behaviors in multiple levels of granularity. They run experiments and show that their clustering method works better than baselines which are k-means and hierarchical clustering algorithm. However, Unlike their work I take temporal variations in the frequency of different behavioral motifs into account by analyzing behavioral motifs as the network is evolving. Furthermore, the networks they consider are commercial large scale social networks that people have been using for a few years before the start of their study. The dataset I using has the communication data between people since the inception of the network. Moreover, in my problem I am pushing limits of sequence analysis in that the events I consider are not diverse by taking into account only joining the network and sending and receiving messages.

Keegan et al. \cite{keegan} demonstrated the potential of mixed methods of analysis for investigating sequence of editing behavior in online knowledge collaborations. They review the literature on online knowledge collaboration and sequence analysis and provide a framework for analysis of sequences in online communities. They argue that analyzing log data using established methods in natural language processing and bioinformatics can provide insights that can answer various kinds of research questions inspired by dynamics of online communities. Moreover, they provide a use case of their framework by analyzing sequence of editing behavior in Wikipedia. They analyze the prevalence and significance of different sequence of editing patterns and empirically assign semantic intuitions to various observed patterns. They capture the strength of sequence methods in analyzing behavioral data and provide a comprehensive framework for formulating similar problems. Unlike my work, they don't specifically consider temporal differences between different events and encode the interval between events by including the temporal data in representation of events. They don't provide a further investigation of how users with different patterns of behavior can be similar or different in the behavior space. 

\subsection{Network Evolution}

There is a large body of research on how social networks form and how they evolve \cite{watts}. Most of these works focus on how the structure of the network changes as the network is evolving. Leskovec et al. \cite{Leskovec:2007:GED:1217299.1217301} show that in a variety of networks having structural and dynamic similarities to social networks, as the number of nodes increase, the number of edges increase superlinearly and the diameter of the network decreases. They show that the graph generator methods didn't capture this property and they provide a new model for graph evolution that correctly incorporates the observed dynamics of evolving networks. Panzarasa et al. \cite{ASI:ASI21015} conduct a study on the college message data set used in the present study  and investigate the degree to which popularity and gregariousness changes among the users as well as structural properties of the network investigated since the inception of the online community under study. 
Unlike previous studies in network evolution, I consider the variations of behavior in fine grained level by investigating the temporal sequences and the correlation of the patterns observed with the social status of the members of the community.

\subsection{User Influence and Centrality}
Many methods and measures have been purposed for estimating a users' influence as well as their centrality in a social network \cite{7364017} \cite{Bakshy:2011:EIQ:1935826.1935845}. Definition of social influence largely varies in research literature depending on the type of social networks, context, and the structural and semantic information considered
(see \cite{Riquelme:2016:MUI:2978438.2978529} for a comprehensive survey on the topic). In this paper, I define social position as users' PageRank centrality \cite{Page98thepagerank} as only structural information is available in the dataset, PageRank is a widely used measure of centrality, and it is relatively an inexpensive to compute. Further, I will use the terms social status, social position, and PageRank of the users interchangeably. 
\subsection{Sequence Analysis}
There are various ways of modelling and analyzing sequence data. RNN-LSTM neural networks \cite{Suhara:2017:DFD:3038912.3052676}, HMM models \cite{Banovic:2016:MUH:2858036.2858557}, and k-gram models \cite{Wang:2016:UCC:2858036.2858107} are among the prominent methods used for modelling sequence data. Following the success of sequence methods in NLP and bio informatics, there has been interest in studying sequence of behavior in recent years (see \cite{Xing:2010:BSS:1882471.1882478} for a brief survey of sequence classification).
In this paper, I adopt k-gram methodology to analyze sequences of user behavior because I believe with observing frequency of recurring sub sequences, i.e. k-grams, of behavior, it's possible to differentiate between individuals with different social positions and predict their PageRank values in the future.
% Your intro should go here. In your report, please include an introduction that introduces and motivates your problem, clearly specifies your research question, and why it is difficult, novel, and interesting. You should have a section devoted to reviewing related work, in the style of the discussion reviews you wrote each week. Then present your data, your methodology, your results, and your interpretation of your results (probably not all in one section). Finally, end with a discussion of your work, what we have learned as a result of your work, and place it 
% in a broader research context, including what follow-up work should be done.

\begin{figure}[t!]
    \centering
    \includegraphics[width= \columnwidth]{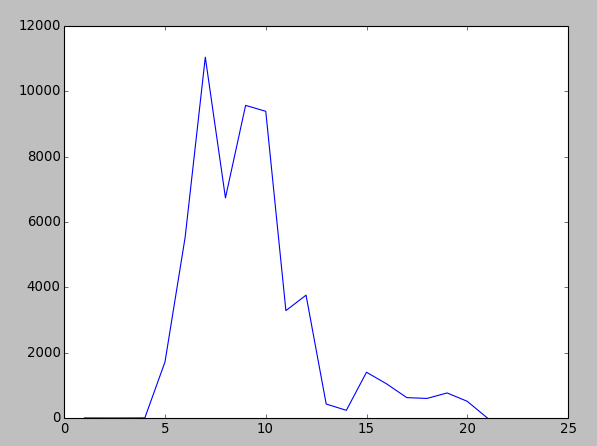}
    \caption{Total amount of weekly messages in the online community}
    \label{activity}
\end{figure}

\section{Dataset}
The dataset used in this study is the time stamped log of messages sent between users in a Facebook-like social network from an online community for students at University of California Irvine \cite{Opsahl_clusteringin}. Every user that sent or received a message is included in the dataset and the total number of users are 1899. Number of messages sent by all the members of the community is 59,835. The data contains the message log between users of the network since the inception of the network and for the duration of more than 6 months. Apart from the message log between members, the time each member joined the network is also recorded in the dataset.
As it can be seen in Figure \ref{activity}, the activity in the network dies after a few weeks. That's why I chose to consider in my analysis only the first twenty weeks in the lifetime of the online community.

\section{Methodology}
\subsection{Networks}
For calculating  and tracking the values of PageRank for the members of the online community through time, I created weekly snapshots of the social network. Every reciprocated message between any two members of the community creates a new undirected tie between those members. Weekly PageRank values are computed for every user that has joined the network before that week utilizing the cumulative network snapshots created. Users that are not in the giant component (most of them have degree of zero) are not taken into consideration when calculating the PageRank values and behavioral sequences. 
\subsection{Behavioral Sequences} \label{bseq}
\begin{figure}[t]
    \centering
    \includegraphics[width= \columnwidth]{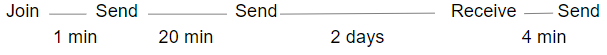}
    \caption{An example of a sequence of user behavior}
    \label{seqlit}
\end{figure}

For each member of the online community behavioral sequences for each week are created. Sequences consist of events that happen with respect to users. There are three event types in my formulation of behavior: joining the network, sending  a message, and receiving a message. In addition to event types, magnitude of time gaps between events are also captured. To make time gaps between the events discrete, time values are replaced with identifiers, representing a range of time values, i.e. buckets. Time gaps are mapped to four buckets:  $<10min, [10min,2hours), [2hours, 1day), \geq 1day$ identified as $A,B,C,D$. For example, the sequence in Figure \ref{seqlit} would be translated into the following sequence:
$J A S B S D R A S$
\subsection{Model} \label{model}
For the problem of predicting social positions of users given their behaviors, PageRank values for a specific target week are computed. Furthermore, the sequence of behavior for a specific duration under study is computed and converted to the representation specified in section \ref{bseq}.
Then k-grams, i.e. frequent sub sequences of behavior with length $k$, and number of occurrence of each k-gram for every user's behavioral sequence are computed. k-gram counts vector for each user is a representation of that user and the counts constitute the features for every user. After defining users in such behavioral space, users are divided to $M$ groups in terms of PageRank values. These groups are created by dividing range of the PageRank values into equidistant bins.
Having represented users as vectors of k-gram counts and belonging to each PageRank group, I consider the PageRank problem as the classifying users into one of $M$ classes of PageRank values given vectors of k-gram counts for representation of user behavior. Note that both behavioral sequences and PageRank values can be computed from different weeks, and I consider both the problem of predicting current PageRank values as well as forecasting the values in the future.

I used random forests classifier for the classification problem specified above and in the next section I will demonstrate that final PageRank values can be predicted with mean absolute error of $0.49$. 

\section{Experiments}
Before moving on to the main contribution of this work, which is forecasting PageRank values, I present several experiments conducted in order to highlighting the correlation of social status with users' behavior in the online community. This experiments underscore the varying nature of social position as well as behavioral traits of different users as the social network is evolving through time, while in the same time portraying that users with different degrees of social position exhibit distinguishable behavioral attributes.

\subsection{Joining Behavior}
Here I investigate how users' final social position, i.e. their PageRank value after the duration of study, is distributed based on their joining date. A large motivating question is whether cumulative advantage is present in determining the value of PageRank and if there's an advantage in joining the network earlier. Because PageRank is essentially a rank and PageRank values form a probability distribution, traditional methods of investigating cumulative advantage that explore connectivity, citations, scientific collaboration, etc. \cite{mattheweffect} can't be readily applied here. Results presented here and in the next sections are consistent with cumulative advantage such that if dynamics governing PageRank followed cumulative advantage, such results would be observed. 
\begin{figure}[t]
    \centering
    \includegraphics[width = \linewidth]{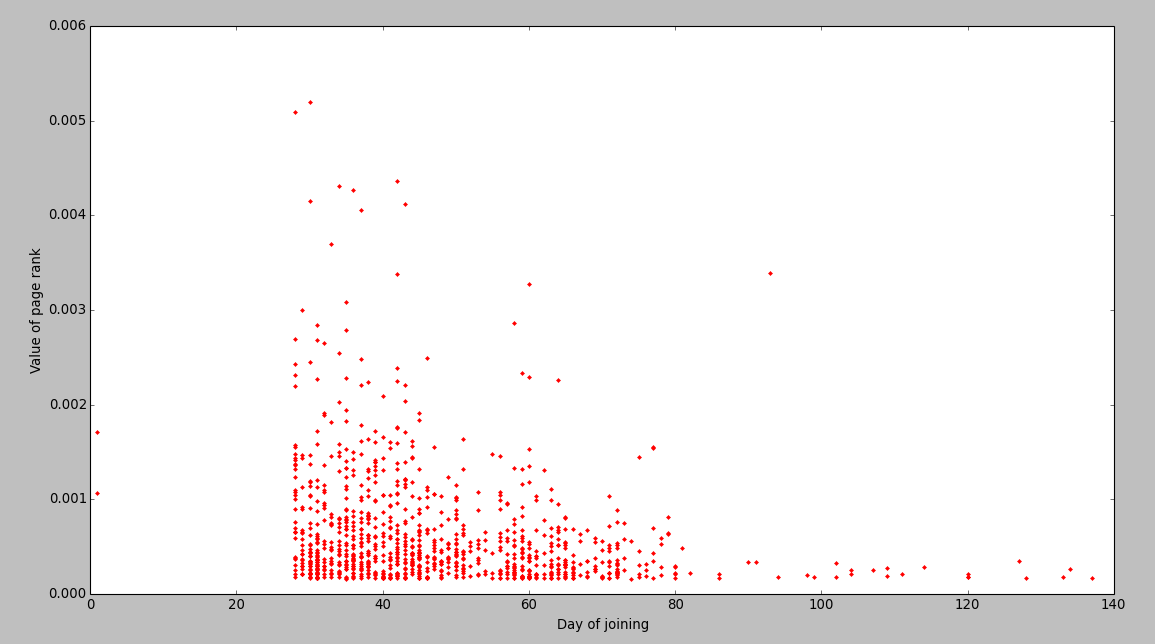}
    \caption{Final PageRank values of users based on their Joining date}
    \label{joinpr}
\end{figure}

As it can be seen in Figure \ref{joinpr}, there's a tendency for people who join early to have higher PageRank values at the end of the duration of study. Spearman correlation is $-0.26$ which indicates a small correlation $(p-value < 0.001)$. This result suggests that there's possibility of presence of cumulative advantage effect in acquiring social position in this online social network. Note that the result here also suggests distinguished joining behavior for people with different levels of social position and helps answering the main research question at least in terms of joining behavior.

\subsection{Receiving New Messages}
In this section, I will investigate distribution of incoming new messages for people with different ranks of social status. Note that in the networks created, incoming new messages are considered as invitations, i.e. friend requests, for creating a new tie. That's why one can argue that the results here represent the probability of people with different social status receiving a request for a new tie. 
\begin{figure}[t]
    \centering
    \includegraphics[width = \linewidth]{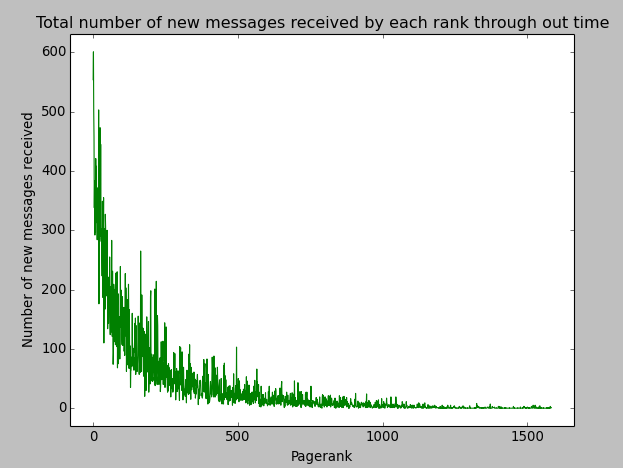}
    \caption{Distribution of total incoming new messages based on the receiver's rank}
    \label{totalmsg}
\end{figure}

As it can be seen in Figure \ref{totalmsg} people that have higher ranks (lower numbers represent higher ranks) tend to receive more new messages. Note that this represents the total number of messages anyone has received and the ranking they had while they received the message. So each point on the plot shows the total number of messages each ranking position receives and since the network is evolving an individual may be ranked differently through time.

\begin{figure}[t]
    \centering
    \includegraphics[width = \linewidth]{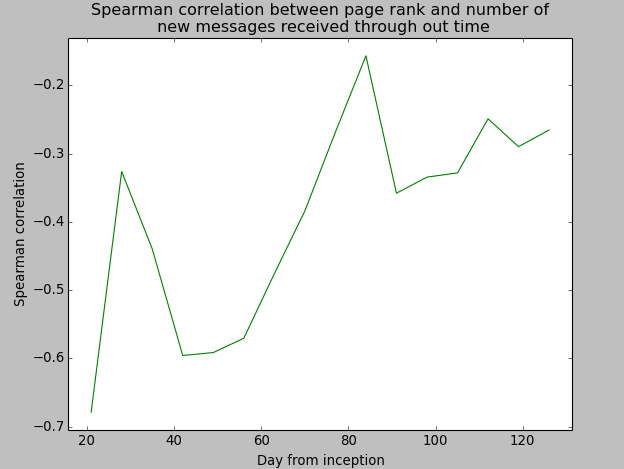}
    \caption{Spearman correlation between weekly PageRank and number of new incoming messages next week}
    \label{spwm}
\end{figure}

Additionally, I calculated the Spearman correlation between weekly social position ranks and the number of new incoming messages for every user that has already joined the network in each particular week. Spearman correlation values are plotted for every week in the duration of study in Figure \ref{spwm}. $p-value<0.001$ for every week.

Both results suggest that users with high PageRank values tend to receive more new messages, i.e. friend requests. Further, users are more likely to message people with better social positions. Similar studies have proven this to be true for other structural measures of social position such as betweenness and closeness centrality \cite{DBLP:journals/corr/abs-1111-6804}.
Moreover, because users that have higher PageRank are receving more new messages, they have better chance of increasing their connectivity and in turn their PageRank. Note that the magnitude of activity of users with different PageRank values is irrelevant in this discussion as I am only considering new messages received, meaning there is no tie between sender and recipient of the messages considered here.

\subsection{Predicting and Forecasting Social Position} \label{main}
In this section, I present the results of predicting PageRank values based on behavioral information available as discussed in section \ref{model}. Thus, having created feature vectors for users representing their behavior in a specific period spanning one or more weeks in the duration of the study, as well as, $M$ groups of PageRank values and determining each user's group, I proceed to predict the group to which a user belongs. I set $M = 7$ for the purpose of this experiments.
Since the problem is classifying on ordinal targets, I measure performance in terms of mean absolute error, i.e. MAE, so that for example, falsely predicting some user to belong to group $i + 1$ instead of their true group $i$ is less punished by the evaluation algorithm than predicting their group to be $i + 5$.

\begin{table}[t]

  \begin{center}
    \begin{tabular}{l|c} 
      \textbf{Model} & \textbf{Mean Absolute Error} \\
      \hline
      Logistic Regression & 0.63\\
      MLP & 0.51\\
      Random Forest & \textbf{0.49}\\
    \end{tabular}
    \caption{Performance of different models on predicting final PageRank given the whole history\label{perf}}
  \end{center}

\end{table}

First, I compare the performance of different classification models for the task of predicting final PageRank value, i.e. PageRank value on week 20, given the whole history of behavioral log from inception of the network until week 20. The performance is represented in terms of mean absolute error obtained after running 8-fold cross validation. The results of this comparison can be seen in Table \ref{perf}. The Random Forest model with $500$ estimators beats the performance of a Multi-layer perceptron model with 5 hidden layers and logistic regression.

This results confirms that it is possible to predict with high accuracy social position given behavioral sequences of members of the online community, proving that people with different levels of social position distinguish themselves in terms of communication behavior. Notice that in many applications $M$ may be chosen to be less than $7$ and accurately predicting users' social positions may not be of concern. To further investigate I also experimented with $M = 3$ simulating a scenario when users are divided to three groups of highly influential, normal, and not influential, divided by their social status. I obtained Mean absolute error $ <0.1 $ across variety of division policies.

I now turn to the problem of forecasting PageRank values. To determine if forecasting social position is possible from information early in the network, I create user behavioral feature vectors by using cumulative information since inception of a network up to a certain week and measure classification performance of the final PageRank values given the cumulative information.
\begin{figure}[t]
    \centering
    \includegraphics[width = \linewidth]{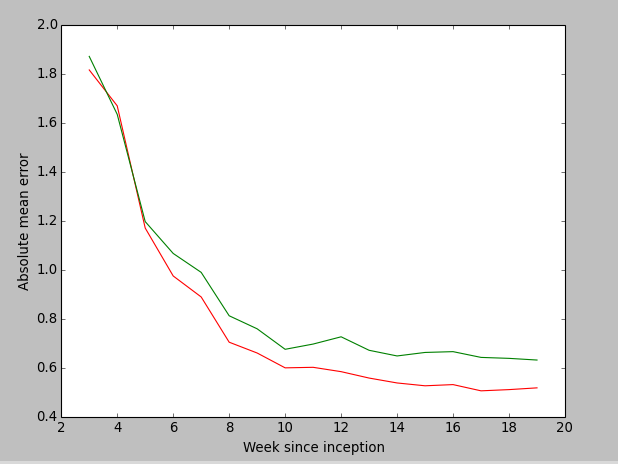}
    \caption{MAE of classification of final PageRank values given distinct information about frequency of different 3-grams until a specific week (red) and MAE of classification of final PageRank values given frequency of all 3-grams until a specific week (green)}
    \label{cper}
\end{figure}

The red plot in Figure \ref{cper} shows mean absolute error of classification with $M = 7$ and $K = 3$ under the random forest model mentioned above. As it can be seen in the plot, the MAE decreases rapidly even when the only information from early in the network's lifetime is available. The large drop in classification error early in time suggests that members communication behavior has high correlation with their social position in the future and that users with different levels of future social position can  distinguish themselves by their early behavior.

To further investigate, the correlation of sequences of behavior with social position, I attempted to predict future PageRank values only exploiting information on frequency of k-grams without differentiating between different behavioral motifs, i.e. feature vectors are a number which is sum of all the elements of behavioral feature vectors used before. This represents a measure of activity  without considering the importance of the order of events. As it can be seen in Figure \ref{cper} (the green plot), by losing information concerning the order of events in behavioral motifs, accuracy of the classification model suffers and MAE increases. However, I expected a larger decrease in accuracy due to importance of order in the behavioral sequences. I believe the reason the order of events are not stressed enough here is the limited scope of the events possible. As you recall, there are only three events in the communication network. I suspect the effects of losing information on sequence order would hurt the accuracy more in an online environment where various types of events and interactions between users are present as it is the case in online social media nowadays \cite{Wang:2016:UCC:2858036.2858107}.

\begin{figure}[t]
    \centering
    \includegraphics[width = \linewidth]{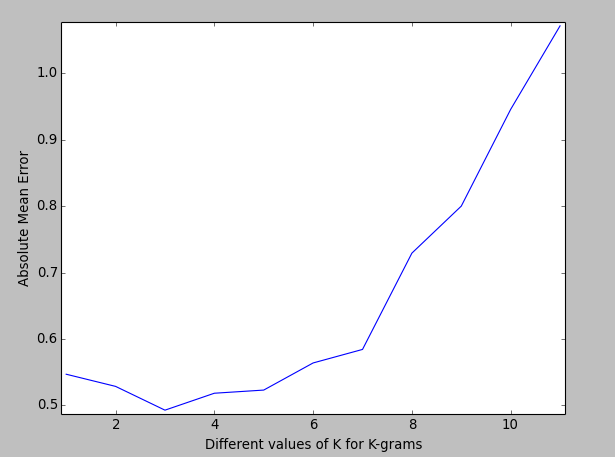}
    \caption{MAE given different values of k as the length of k-grams, i.e. behavioral motifs, for the case of predicting PageRank values}
    \label{k}
\end{figure}
To further study the power of sequence models, I modified the length of behavioral motifs. Figure \ref{k} shows the mean absolute error for different values of $k$, i.e. the length of behavioral motifs. The optimal value of $k$ is $3$. As $k$ gets smaller the focus on sequential nature of the data is reduced. For instance, when $k = 1$ the model only considers the frequency of sending and receiving messages. This confirms my previous result, by stressing the importance of order in sequences rather than only considering the magnitude of activity for distinguishing users with different social positions. On the other hand, as $k$ gets larger the frequency of motifs that represent a certain behavioral trait decreases, the variance in data increases, and the performance of the model decreases.

\begin{figure}[t]
    \centering
    \includegraphics[width = \linewidth]{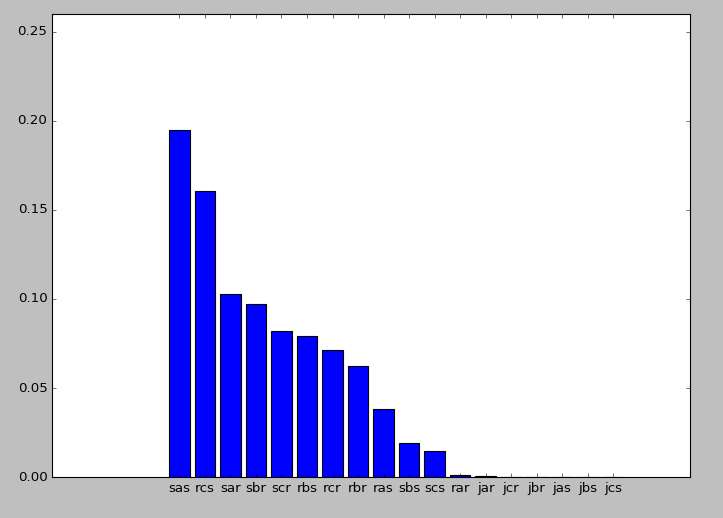}
    \caption{Mean increase (percentage) in error as a result of perturbing the frequency of each feature}
    \label{freqcorr}
\end{figure}
Lastly, I studied the correlations of different k-grams (with $k = 3$) with social position. By identifying the most important 3-grams in that provide the most information when predicting social position. A popular method for analyzing feature importance is to directly measure the impact of each feature on error or accuracy of the classification algorithm. The idea is to perturb the values of every feature and measure the degree to which this permutation decreases the accuracy of the model. The features are the frequency of occurrence of each behavioral motif. Thus, the extent to which perturbing a feature will increase the error in classifying the social position of individuals is highly representative of how that feature can describe the relationship between user behavior and social position. I perturbed the values of each feature by shuffling the values for that feature in the test set and computed the mean increase in mean absolute error across $25$ random splits of training and test data for predicting PageRank values based on the whole observed history via the best performing random forest algorithm discussed in section \ref{main}. As it has been illustrated in Figure \ref{freqcorr}, some behavioral features such as $SAS$ and $RCS$ are highly indicative of users' social position whereas some are not indicative at all (e.g. sequences starting with $J$ are not indicative because everyone in the network joins once and the sequence observed only once is of little value). Interestingly, performance of the algorithm does not suffer much if a few variables are perturbed. For instance, It can be seen that perturbing the most indicative feature increases the error by $20\%$ (which is a small number considering the classification error for 8-fold cross validation was $0.49$). I suspect the reason is that all features are contributing to identifying users with different levels of PageRank and we cannot point to a small set of features (3-gram frequencies) to explain user behavior and PageRank. In fact, this confirms my intuition that behavior expressed in terms of frequency of various behavioral motifs can better describe user social position rather than observing a specific behavior or aggregation of behavioral statistics.

\section{Discussion}
In this paper, I set out to discover if a user's behavior in a communication network has correlation with their social position. Behavior was considered as the number of frequent behavioral motifs and social position was represented as users' PageRank values on networks of communication constructed. Three important discoveries were made. \newline
First, I discovered that users with high social positions have a better chance of maintaining and increasing their social positions. Users who join early and have spent more time in the network can tend to have a better social position and users that have better social positions received larger number of incoming new messages, i.e. friend requests, that will enable them to increase their connectivity and in turn, their social position. Since PageRank is essentially a ranking between users and a shared resource between members of a community, traditional definitions of cumulative advantage cannot be applied in this case. Even so, the observations are consistent with effects of cumulative advantage.

Second, users with various levels of social status are distinguishable by their behavior. In fact, the most important discovery of this work is that users' behavioral traits are a good indication of  their future social status in the network. I have shown that it is possible to classify users in different social position groups in the future based on their behavior early in the online community's lifetime. I believe this finding can lead to better understanding of  users' behavior and how dynamics of online behavior can relate to  influence or prominence among members of an online communities. In larger context, the applications of identifying influential users in a community are abundant. Identifying key players in a market, leaders of a criminal network, and eminent members of a political or academic organization are a few examples of such applications. Additionally, PageRank is a widely used method of determining importance of entities in a networks prevalent in nature and human artifacts. Thus, The observations here may also hold true outside of the social context, for instance, by considering interaction between entities in biological networks, webpages, cellular networks and internet, etc.

Finally, I have highlighted the importance of order and time in analyzing user behavior. It is shown that social position can be better predicted by taking into account the sequential and temporal nature of the data. 
Recently, there has been a large interest in the research community for proposing methods that maximize the amount of information captured in the modelling of sequence date. Following the success of sequence methods in bioinformatics and NLP, here I have presented the power of sequence methods in capturing behavioral information. As I have shown here, by omitting the sequential nature of the data in modelling of a problem, useful information may be lost which may reduce the effectiveness of the methods used.

\section{Limitations and Future Work}

In this paper, I have shown that the by utilizing the power of sequence modelling, user behavior  and social position in online communities can be better understood, however, the proposed methodology has some limitations. 

Activity in the online social network under study dies after twenty weeks which limits the duration of study and stability of the results for longer periods. In addition, the largest amount of activity happens in the first weeks since the online community is established and one cannot deduce whether the observations presented here, were mostly due to high activity or the amount of time passed since network's inception. Even so, the results are still significant by demonstrating how early behavior can convey information about future social position in a community. However, investigation on networks that reach a level of stability in activity magnitude and studies over longer periods are required to validate the results of this study in the general case.

Furthermore, there are more powerful methods of modelling sequences that can represent longer time dependencies and directly incorporate the temporal and sequential nature of log data. The focus here was to show the importance of behavior sequences by using k-gram models, however, utilizing more powerful methods of sequence modelling may lead to more accurate representation of users in terms of their communication behavior. Exploration of these methods are left for future research. 

Lastly, the observations of this study are compatible with the effects of cumulative advantage for social position in a social network. However, since traditional definitions of cumulative advantage cannot be applied to variation in users' PageRank values, I have provided indirect evidence of a presence of a dynamic that enables higher ranked users to have a better chance of maintaining and increasing their status in the online community. Further empirical investigations of how PageRank and in general social status vary through time for evolving networks comprised of members with different levels of social status are required to better explain the results obtained here.

\bibliography{SSCB}

\begin{thebibliography}{30}
\expandafter\ifx\csname natexlab\endcsname\relax\def\natexlab#1{#1}\fi

\bibitem[{Abbasi et~al.(2011)Abbasi, Hossain, and
  Leydesdorff}]{DBLP:journals/corr/abs-1111-6804}
Alireza Abbasi, Liaquat Hossain, and Loet Leydesdorff. 2011.
\newblock \href {http://arxiv.org/abs/1111.6804} {Betweenness centrality as a
  driver of preferential attachment in the evolution of research collaboration
  networks}.
\newblock \emph{CoRR}, abs/1111.6804.

\bibitem[{Anderson et~al.(2012)Anderson, Huttenlocher, Kleinberg, and
  Leskovec}]{Anderson12}
Ashton Anderson, Daniel Huttenlocher, Jon Kleinberg, and Jure Leskovec. 2012.
\newblock \href {https://doi.org/10.1145/2124295.2124378} {Effects of user
  similarity in social media}.
\newblock In \emph{Proceedings of the Fifth ACM International Conference on Web
  Search and Data Mining}, WSDM '12, pages 703--712, New York, NY, USA. ACM.

\bibitem[{Bakshy et~al.(2011)Bakshy, Hofman, Mason, and
  Watts}]{Bakshy:2011:EIQ:1935826.1935845}
Eytan Bakshy, Jake~M. Hofman, Winter~A. Mason, and Duncan~J. Watts. 2011.
\newblock \href {https://doi.org/10.1145/1935826.1935845} {Everyone's an
  influencer: Quantifying influence on twitter}.
\newblock In \emph{Proceedings of the Fourth ACM International Conference on
  Web Search and Data Mining}, WSDM '11, pages 65--74, New York, NY, USA. ACM.

\bibitem[{Banovic et~al.(2016)Banovic, Buzali, Chevalier, Mankoff, and
  Dey}]{Banovic:2016:MUH:2858036.2858557}
Nikola Banovic, Tofi Buzali, Fanny Chevalier, Jennifer Mankoff, and Anind~K.
  Dey. 2016.
\newblock \href {https://doi.org/10.1145/2858036.2858557} {Modeling and
  understanding human routine behavior}.
\newblock In \emph{Proceedings of the 2016 CHI Conference on Human Factors in
  Computing Systems}, CHI '16, pages 248--260, New York, NY, USA. ACM.

\bibitem[{Benevenuto et~al.(2009)Benevenuto, Rodrigues, Cha, and
  Almeida}]{Benevenuto:2009:CUB:1644893.1644900}
Fabr\'{\i}cio Benevenuto, Tiago Rodrigues, Meeyoung Cha, and Virg\'{\i}lio
  Almeida. 2009.
\newblock \href {https://doi.org/10.1145/1644893.1644900} {Characterizing user
  behavior in online social networks}.
\newblock In \emph{Proceedings of the 9th ACM SIGCOMM Conference on Internet
  Measurement}, IMC '09, pages 49--62, New York, NY, USA. ACM.

\bibitem[{Chu et~al.(2012)Chu, Gianvecchio, Wang, and Jajodia}]{6280553}
Z.~Chu, S.~Gianvecchio, H.~Wang, and S.~Jajodia. 2012.
\newblock \href {https://doi.org/10.1109/TDSC.2012.75} {Detecting automation of
  twitter accounts: Are you a human, bot, or cyborg?}
\newblock \emph{IEEE Transactions on Dependable and Secure Computing},
  9(6):811--824.

\bibitem[{De~Choudhury et~al.(2013)De~Choudhury, Gamon, Counts, and
  Horvitz}]{predicting-depression-via-social-media}
Munmun De~Choudhury, Michael Gamon, Scott Counts, and Eric Horvitz. 2013.
\newblock \href
  {https://www.microsoft.com/en-us/research/publication/predicting-depression-via-social-media/}
  {Predicting depression via social media}.
\newblock AAAI.

\bibitem[{Eagle and Pentland(2009)}]{Eagle09eigenbehaviors:identifying}
Nathan Eagle and Alex~(Sandy) Pentland. 2009.
\newblock Eigenbehaviors: identifying structure in routine.

\bibitem[{Hu et~al.(2007)Hu, Zeng, Li, Niu, and
  Chen}]{Hu:2007:DPB:1242572.1242594}
Jian Hu, Hua-Jun Zeng, Hua Li, Cheng Niu, and Zheng Chen. 2007.
\newblock \href {https://doi.org/10.1145/1242572.1242594} {Demographic
  prediction based on user's browsing behavior}.
\newblock In \emph{Proceedings of the 16th International Conference on World
  Wide Web}, WWW '07, pages 151--160, New York, NY, USA. ACM.

\bibitem[{Jin et~al.(2013)Jin, Chen, Wang, Hui, and Vasilakos}]{6588663}
L.~Jin, Y.~Chen, T.~Wang, P.~Hui, and A.~V. Vasilakos. 2013.
\newblock \href {https://doi.org/10.1109/MCOM.2013.6588663} {Understanding user
  behavior in online social networks: a survey}.
\newblock \emph{IEEE Communications Magazine}, 51(9):144--150.

\bibitem[{Jurgens and Lu(2012)}]{jurgens}
David Jurgens and Tsai-Ching Lu. 2012.
\newblock Temporal motifs reveal the dynamics of editor interactions in
  wikipedia.

\bibitem[{Keegan et~al.(2015)Keegan, Lev, and Arazy}]{keegan}
Brian~C. Keegan, Shakked Lev, and Ofer Arazy. 2015.
\newblock \href {http://arxiv.org/abs/1508.04819} {Analyzing organizational
  routines in online knowledge collaborations: {A} case for sequence analysis
  in {CSCW}}.
\newblock \emph{CoRR}, abs/1508.04819.

\bibitem[{Kossinets and Watts(2006)}]{watts}
Gueorgi Kossinets and Duncan Watts. 2006.
\newblock Empirical analysis of an evolving social network.
\newblock 311:88--90.

\bibitem[{Leskovec et~al.(2007)Leskovec, Kleinberg, and
  Faloutsos}]{Leskovec:2007:GED:1217299.1217301}
Jure Leskovec, Jon Kleinberg, and Christos Faloutsos. 2007.
\newblock \href {https://doi.org/10.1145/1217299.1217301} {Graph evolution:
  Densification and shrinking diameters}.
\newblock \emph{ACM Trans. Knowl. Discov. Data}, 1(1).

\bibitem[{Mahmud et~al.(2014)Mahmud, Nichols, and
  Drews}]{Mahmud:2014:HLI:2648782.2528548}
Jalal Mahmud, Jeffrey Nichols, and Clemens Drews. 2014.
\newblock \href {https://doi.org/10.1145/2528548} {Home location identification
  of twitter users}.
\newblock \emph{ACM Trans. Intell. Syst. Technol.}, 5(3):47:1--47:21.

\bibitem[{Mislove et~al.(2010)Mislove, Viswanath, Gummadi, and
  Druschel}]{Mislove:2010:YYK:1718487.1718519}
Alan Mislove, Bimal Viswanath, Krishna~P. Gummadi, and Peter Druschel. 2010.
\newblock \href {https://doi.org/10.1145/1718487.1718519} {You are who you
  know: Inferring user profiles in online social networks}.
\newblock In \emph{Proceedings of the Third ACM International Conference on Web
  Search and Data Mining}, WSDM '10, pages 251--260, New York, NY, USA. ACM.

\bibitem[{Opsahl and Panzarasa()}]{Opsahl_clusteringin}
Tore Opsahl and Pietro Panzarasa.
\newblock Clustering in weighted networks.
\newblock \emph{Social Networks}, page 2009.

\bibitem[{Page et~al.(1998)Page, Brin, Motwani, and
  Winograd}]{Page98thepagerank}
Larry Page, Sergey Brin, R.~Motwani, and T.~Winograd. 1998.
\newblock The pagerank citation ranking: Bringing order to the web.

\bibitem[{Panzarasa et~al.(2009)Panzarasa, Opsahl, and Carley}]{ASI:ASI21015}
Pietro Panzarasa, Tore Opsahl, and Kathleen~M. Carley. 2009.
\newblock \href {https://doi.org/10.1002/asi.21015} {Patterns and dynamics of
  users' behavior and interaction: Network analysis of an online community}.
\newblock \emph{Journal of the American Society for Information Science and
  Technology}, 60(5):911--932.

\bibitem[{Perc(2014)}]{mattheweffect}
Matjaž Perc. 2014.
\newblock The matthew effect in empirical data.
\newblock 11.

\bibitem[{Rao et~al.(2015)Rao, Spasojevic, Li, and Dsouza}]{7364017}
A.~Rao, N.~Spasojevic, Z.~Li, and T.~Dsouza. 2015.
\newblock \href {https://doi.org/10.1109/BigData.2015.7364017} {Klout score:
  Measuring influence across multiple social networks}.
\newblock In \emph{2015 IEEE International Conference on Big Data (Big Data)},
  pages 2282--2289.

\bibitem[{Riquelme and
  Gonz\'{a}lez-Cantergiani(2016)}]{Riquelme:2016:MUI:2978438.2978529}
Fabi\'{a}n Riquelme and Pablo Gonz\'{a}lez-Cantergiani. 2016.
\newblock \href {https://doi.org/10.1016/j.ipm.2016.04.003} {Measuring user
  influence on twitter}.
\newblock \emph{Inf. Process. Manage.}, 52(5):949--975.

\bibitem[{Suhara et~al.(2017)Suhara, Xu, and
  Pentland}]{Suhara:2017:DFD:3038912.3052676}
Yoshihiko Suhara, Yinzhan Xu, and Alex~'Sandy' Pentland. 2017.
\newblock \href {https://doi.org/10.1145/3038912.3052676} {Deepmood:
  Forecasting depressed mood based on self-reported histories via recurrent
  neural networks}.
\newblock In \emph{Proceedings of the 26th International Conference on World
  Wide Web}, WWW '17, pages 715--724, Republic and Canton of Geneva,
  Switzerland. International World Wide Web Conferences Steering Committee.

\bibitem[{Tang and Yang(2010)}]{5484779}
Xuning Tang and C.~C. Yang. 2010.
\newblock \href {https://doi.org/10.1109/ISI.2010.5484779} {Identifing
  influential users in an online healthcare social network}.
\newblock In \emph{2010 IEEE International Conference on Intelligence and
  Security Informatics}, pages 43--48.

\bibitem[{Van Der~Aalst et~al.(2005)Van Der~Aalst, Reijers, and
  Song}]{VanDerAalst:2005:DSN:1107360.1107389}
Wil M.~P. Van Der~Aalst, Hajo~A. Reijers, and Minseok Song. 2005.
\newblock \href {https://doi.org/10.1007/s10606-005-9005-9} {Discovering social
  networks from event logs}.
\newblock \emph{Comput. Supported Coop. Work}, 14(6):549--593.

\bibitem[{Varol et~al.(2017)Varol, Ferrara, Davis, Menczer, and
  Flammini}]{DBLP:journals/corr/VarolFDMF17}
Onur Varol, Emilio Ferrara, Clayton~A. Davis, Filippo Menczer, and Alessandro
  Flammini. 2017.
\newblock \href {http://arxiv.org/abs/1703.03107} {Online human-bot
  interactions: Detection, estimation, and characterization}.
\newblock \emph{CoRR}, abs/1703.03107.

\bibitem[{Wang et~al.(2016)Wang, Zhang, Tang, Zheng, and
  Zhao}]{Wang:2016:UCC:2858036.2858107}
Gang Wang, Xinyi Zhang, Shiliang Tang, Haitao Zheng, and Ben~Y. Zhao. 2016.
\newblock \href {https://doi.org/10.1145/2858036.2858107} {Unsupervised
  clickstream clustering for user behavior analysis}.
\newblock In \emph{Proceedings of the 2016 CHI Conference on Human Factors in
  Computing Systems}, CHI '16, pages 225--236, New York, NY, USA. ACM.

\bibitem[{Wang et~al.(2015)Wang, Yuan, Lian, Xu, Xie, Chen, and Rui}]{wang}
Yingzi Wang, Nicholas~Jing Yuan, Defu Lian, Linli Xu, Xing Xie, Enhong Chen,
  and Yong Rui. 2015.
\newblock \href {https://doi.org/10.1145/2783258.2783350} {Regularity and
  conformity: Location prediction using heterogeneous mobility data}.
\newblock In \emph{Proceedings of the 21th ACM SIGKDD International Conference
  on Knowledge Discovery and Data Mining}, KDD '15, pages 1275--1284, New York,
  NY, USA. ACM.

\bibitem[{Xing et~al.(2010)Xing, Pei, and
  Keogh}]{Xing:2010:BSS:1882471.1882478}
Zhengzheng Xing, Jian Pei, and Eamonn Keogh. 2010.
\newblock \href {https://doi.org/10.1145/1882471.1882478} {A brief survey on
  sequence classification}.
\newblock \emph{SIGKDD Explor. Newsl.}, 12(1):40--48.

\bibitem[{Zhong et~al.(2015)Zhong, Yuan, Zhong, Zhang, and
  Xie}]{Zhong:2015:YYG:2684822.2685287}
Yuan Zhong, Nicholas~Jing Yuan, Wen Zhong, Fuzheng Zhang, and Xing Xie. 2015.
\newblock \href {https://doi.org/10.1145/2684822.2685287} {You are where you
  go: Inferring demographic attributes from location check-ins}.
\newblock In \emph{Proceedings of the Eighth ACM International Conference on
  Web Search and Data Mining}, WSDM '15, pages 295--304, New York, NY, USA.
  ACM.

\end{thebibliography}
\bibliographystyle{SSCB}

\appendix

\end{document}